\documentclass[%
reprint,
superscriptaddress,
amsmath,amssymb,
aps,
prl,
sort&compress,numbers,super
]{revtex4-2}

\usepackage{graphicx}
\usepackage{dcolumn}
\usepackage{bm}
\usepackage{soul}
\usepackage{xcolor}

\usepackage[caption=false]{subfig}

\begin{document}

\title{Proximity-Induced Exchange Interaction: a New Pathway for Quantum Sensing using Spin Centers in Hexagonal Boron Nitride}

\author{Lingnan Shen}
    \affiliation{Department of Physics, University of Washington, Seattle, WA, USA}
\author{Di Xiao}
    \email{dixiao@uw.edu}
    \affiliation{Department of Material Science and Engineering, University of Washington, Seattle, WA, USA}
    \affiliation{Department of Physics, University of Washington, Seattle, WA, USA}
    \affiliation{Pacific Northwest National Laboratory, Richland, WA, USA}
\author{Ting Cao}
    \email{tingcao@uw.edu}
    \affiliation{Department of Material Science and Engineering, University of Washington, Seattle, WA, USA}

\date{\today}

\begin{abstract}
Defects in hexagonal boron nitride (hBN), a two-dimensional van der Waals material, have 
raised wide range interest for its potential in various quantum applications. 
Due to hBN's 2D nature, spin center in hBN can be engineered in close proximity to target material, 
providing advantages over their 3D counterparts, such as nitrogen-vacancy (NV) center in diamond. 
Here we propose a novel quantum sensing protocol driven by exchange interaction between spin
center in hBN and the underlying magnetic substrate induced by magnetic proximity effect. 
By first-principle calculation, we
demonstrate the induced exchange interaction dominates over dipole-dipole interaction
by orders of magnitude when in proximity. The interaction remains antiferromagnetic across all stacking
configuration between the spin center in hBN and the target van der Waals magnets.
Additionally, we explored the scaling behavior of the exchange field as a function of the spatial separation between the spin center and the targets.

\end{abstract}

\maketitle

\begin{figure}
    \centering
    \textbf{For Table of Contents Only}
    \includegraphics[width=0.75\linewidth]{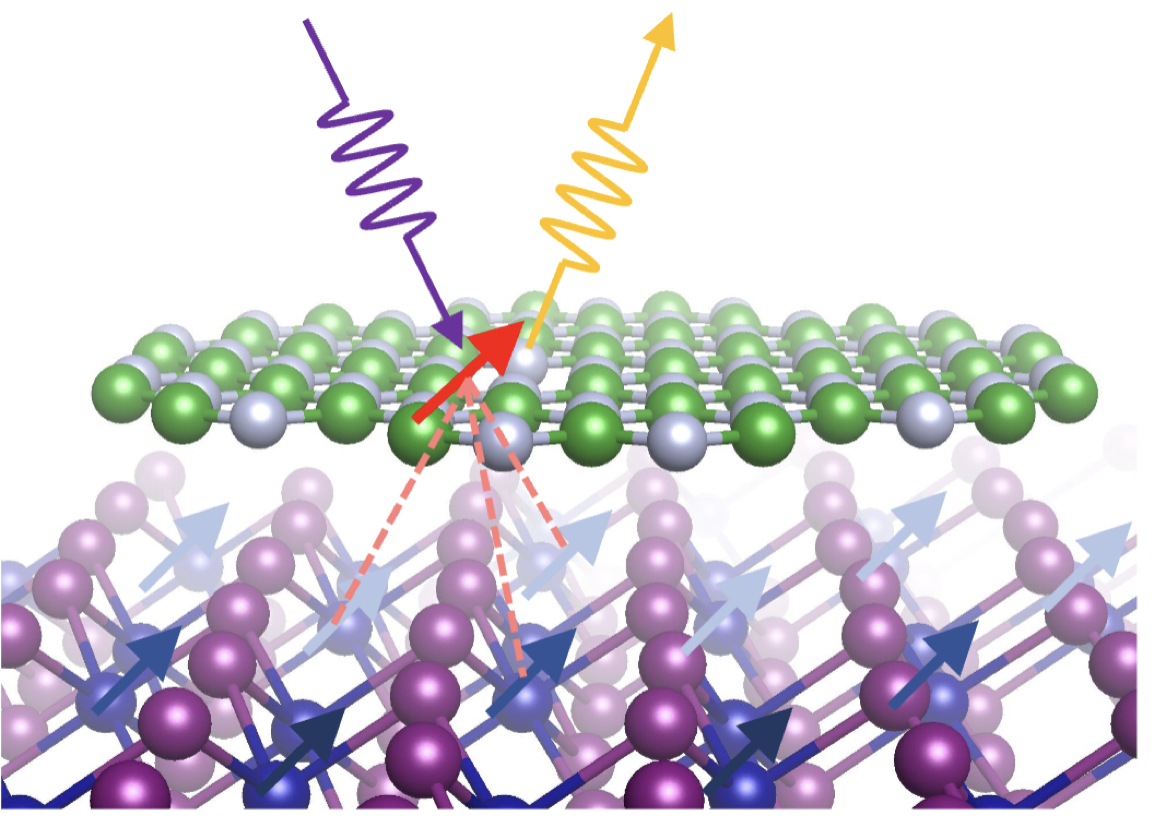}
\end{figure}
Quantum sensing based on solid-state platform has been successful in delivering
high resolution and stable measurements of various physical quantities such as temperature, pressure, strain, magnetic, electric, and even gravitational fields
\cite{degen_quantum_2017, schirhagl_nitrogen-vacancy_2014, casola_probing_2018, wolfowicz_quantum_2021}.
High-precision, high-spatial resolution detection of magnetic fields is particularly important because it enables detailed understanding of physical phenomena ranging from fundamental quantum mechanics to many intricate biological processes
\cite{song_direct_2021, finco_imaging_2021, huang_revealing_2023, barry_optical_2016, aslam_quantum_2023}.
Among the candidate platforms for probing magnetic field, 
color centers with particular spin and optical interface suitable
for manipulation have been extensively investigated
\cite{koehl_room_2011, zhang_material_2020, freysoldt_first-principles_2014}.
Current sensing applications predominantly rely on Zeeman interaction between the color center 
spins and the small stray field generated by the sensing target. 
Such Zeeman splitting is in the microwave range and usually measured by optically
detected magnetic resonance (ODMR).

One widely used color center is the NV center in diamond \cite{song_direct_2021, maze_nanoscale_2008, kolkowitz_probing_2015, steinert_magnetic_2013}.
Despite its success, the NV center suffers from several intrinsic
limitations. 
Firstly, high quality NV centers are usually embedded in the diamond bulk, 
as it is difficult to obtain NV center with long coherence
time near diamond surface due to the noise from surface dangling bonds or loss of desirable charge state 
\cite{romach_spectroscopy_2015, bluvstein_identifying_2019}. 
This bulk embedment simultaneously 
impedes the ability to probe the sensing target at extreme proximity for atomic resolution.
Furthermore, the magnetic sensing capabilities of NV centers are constrained by their inability to 
detect the magnetism of underlying targets with net zero magnetic moments, 
like antiferromagnetic (AFM) materials, which results in a vanishing stray field signal,
excluding their use as \textit{in situ} sensors for such applications.
Secondly, Zeeman interaction splitting is determined
by the projection of stray field to the quantization axis of color center. 
If the stray field is orthogonal
to the quantization axis, there will be no signal on the ODMR spectrum.
Thus, single NV center is only sensitive to the variation of stray field along the 
pre-determined quantization axis.
These limitations mentioned above are generic for color centers embedded in bulk semiconductors.

A material platform that supports a fundamentally new sensing paradigm could be provided by 
defects in hexagonal boron nitride (hBN) \cite{attaccalite_efficient_2013, durand_optically_2023, 
gottscholl_room_2021, gottscholl_initialization_2020, chen_photophysical_2021,
mathur_excited-state_2022, healey_quantum_2023, huang_wide_2022, gottscholl_spin_2021}
, a two-dimensional (2D) van der Waals (vdW) material. 
2D vdW materials can be engineered to atomically thin layer while
free from dangling bonds \cite{dean_boron_2010, radisavljevic_single-layer_2011}.
This effectively resolves the two limitations we encounter with color centers in bulk semiconductors.
The integration of 2D vdW materials into heterostructures facilitates the engineering of 
defects within a few layers from the interface with target sample, providing an opportunity for a new 
paradigm of ultrasensitive, \textit{in situ} quantum sensing \cite{novoselov_2d_2016}.

\begin{figure}[h]
    \subfloat[\label{fig:schematic_afm}]{%
    \includegraphics[width=0.45\linewidth]{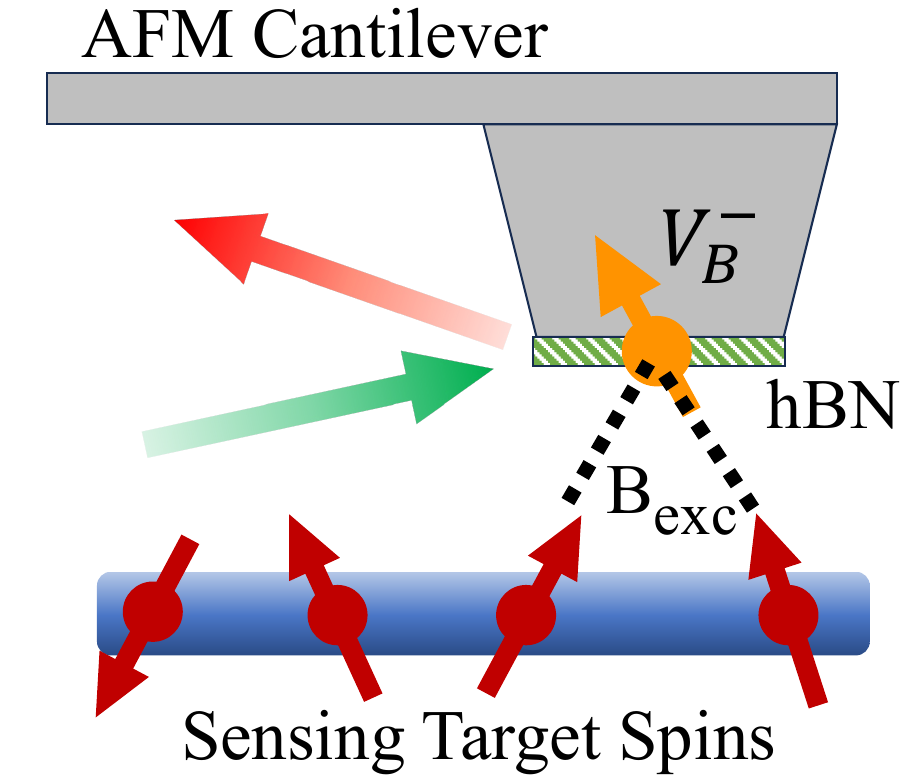}%
    }
    \hfill
    \subfloat[\label{fig:schematic_vdw}]{%
    \includegraphics[width=0.4\linewidth]{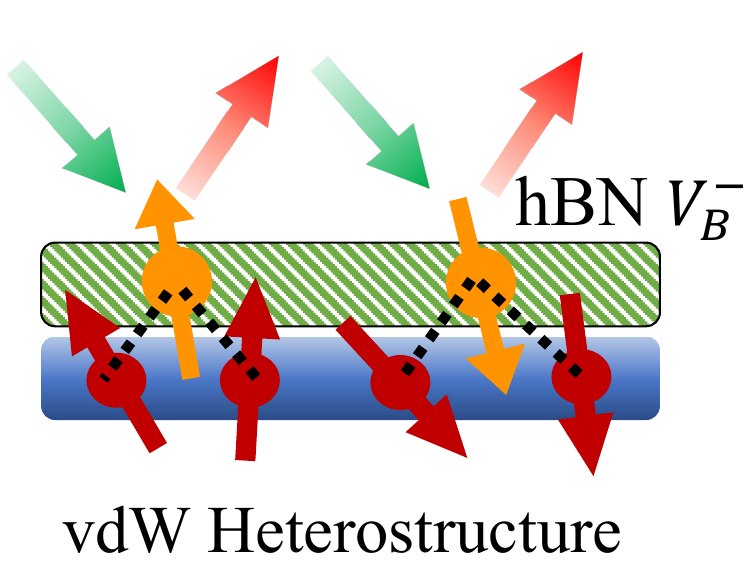}%
    }
    \hfill
    \subfloat[\label{fig:schematic_bandalignment}]{%
    \includegraphics[width=0.55\linewidth]{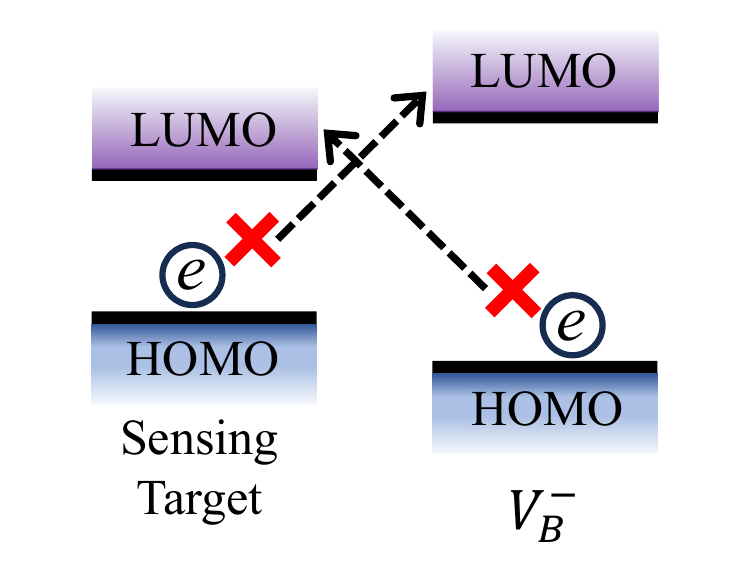}%
    }
    \caption{\label{fig:schematic} Schematic of quantum
    sensing based on proximity-induced exchange interactions. Orange and dark red arrow denote the
    magnetic moments of $V_B^-$ spin center and the sensing target (blue layer), respectively. 
    Optical manipulation and readout of $V_B^-$ spin state are marked with light green and red arrow.
    (a) hBN layer containing a $V_B^-$ spin center coated on atomic force microscopy cantilever
    and positioned in close proximity to the sensing target.
    (b) vdW heterostructure formed by hBN containing ensembles of $V_B^-$ spin center 
    and magnetic material, i.e., the sensing target.
    Dashed lines denote strong proximity induced exchange interactions that
    couple the $V_B^-$ spin centers with local magnetic environment.
    (c) Schematic of possible band alignment scenario for $V_B^-$ center and
    the sensing target. Such alignment prevents charge transfer between $V_B^-$ center and
    the sensing target, ensuring the electronic stability of $V_B^-$ center for
    quantum sensing application.}
\end{figure}

In this letter, we propose a novel quantum sensing protocol driven by the exchange interaction between
spin center in 2D vdW material and the target sample, demonstrated in Fig. \ref{fig:schematic}. 
Our \textit{ab-initio} calculations demonstrate a gigantic
exchange interaction on the order of meV between the negatively charged boron-vacancy ($V_B^-$) center 
and a magnetic substrate when engineered to proximity in a heterostructure.
Remarkably, the strength of such exchange interaction remains robust across all stacking configurations,
and dominates over classical dipole-dipole or stray-field interactions.
Our proposal addresses obstacles encountered with stray-field based sensing
protocol.
This work thus introduces a new exchange interaction-driven \textit{in situ} quantum sensing 
scheme with potential for ultrahigh sensitivity.

\section{Results and Discussion}
\subsection{Sensing Design}
hBN hosts a variety of optically addressable defects that remain robust at room temperature and 
pressure \cite{attaccalite_efficient_2013, stern_room-temperature_2022, li_identification_2022, guo_generation_2022, jin_fabrication_2009, mendelson_identifying_2021, chen_photophysical_2021, chejanovsky_single-spin_2021}.
The negatively charged boron-vacancy ($V_B^-$) in hBN attracts particular interest
due to its maturity in fabrication process and depth of research into its electronic and optical properties.
Many recent studies have demonstrated the ability to initialize, manipulate, and readout
the spin state of $V_B^-$ \cite{gottscholl_room_2021, gottscholl_initialization_2020, 
mathur_excited-state_2022, liu_coherent_2022, gao_nuclear_2022}.
Since its first experimental report, $V_B^-$ quantum sensing application has spanned among
static magnetic field, temperature, strain, pressure, and spin fluctuation \cite{healey_quantum_2023, 
huang_wide_2022, gottscholl_spin_2021, liu_temperature-dependent_2021, lyu_strain_2022, yang_spin_2022}. In this work, we will primarily focus on $V_B^-$ as the quantum sensors.

We start by outlining two general design principles of quantum sensing based on 
proximity-induced exchange interactions.
First, the quantum sensor, i.e., the spin center, has to be structurally and electronically stable.
This requires no covalent or ionic chemical bonds, or charge transfer between 
the desirable spin center and the sensing target, shown in Fig. \ref{fig:schematic_bandalignment}. 
To this end, we identified several 2D magnetic semiconductors, such as CrI$_3$, $\text{CrCl}_3$, and CrSBr,
as the sensing target of current interests, due to their 
technical importance \cite{lu_meron-like_2020, huang_layer-dependent_2017, cai_atomically_2019, lee_magnetic_2021, telford_layered_2020} and compatibility with the $V_B^-$ center.

The second design principle of proximity quantum sensing is a significant exchange interaction 
between the quantum sensor and target, which causes measurable changes in the 
electronic structures of the quantum sensors.
The $V_B^-$ center features a spin triplet ground state with a total magnetic moment of $2\mu_B$. 
The exchange interaction will split the $m_s = +1$ and $m_s = -1$ spin states within the $V_B^-$ ground state manifold, which could be measured by experiment such as
optically detected magnetic resonance.
Upon forming a vdW interface, we expect the exchange interaction between the $V_B^-$ and the 2D magnetic semiconductors, such 
as CrI$_3$, to be comparable in size to the pairwise exchange interactions between adjacent magnet layers. 
In addition, the robustness of our proposed sensing scheme is ensured by the dominance of such significant exchange interaction over other interactions such as dipole-dipole interaction.

Potential experimental realizations of this novel quantum sensing protocol
include direct integration of
the $V_B^-$ spin center with an atomic force microscopy cantilever to
scan the sensing target at proximity, as shown in Fig. \ref{fig:schematic_afm}. The hBN sheets containing $V_B^-$ spin center
could be coated onto the surface of atomic force microscopy cantilever.
Alternatively, one can incorporate $V_B^-$ center and the sensing target into a vdW heterostructure,
depicted in Fig. \ref{fig:schematic_vdw},
with existing technology such as optically detected magnetic resonance (ODMR) for 
measurement purpose.
The inherent 2D nature adds extra flexibility in experimental setup, allowing for the adaptation of the hBN sheet onto various surfaces.

\begin{figure*}
  \centering
  \subfloat[hBN $V_B$ defect.\label{fig:band_align_vb}]{%
    \includegraphics[width=0.218\linewidth]{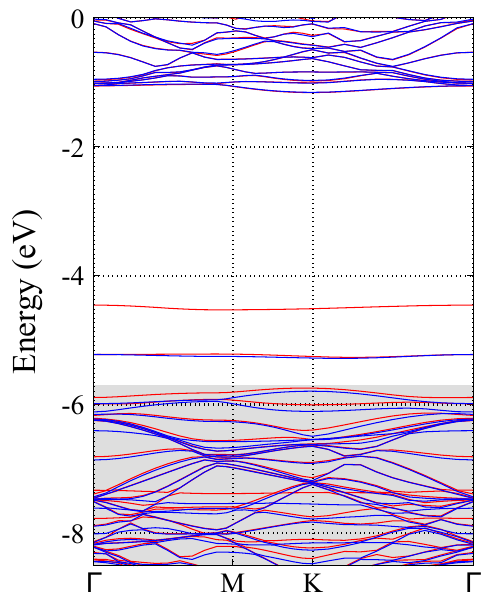}%
  }
  \hfill
  \subfloat[CrI$_3$.\label{fig:band_align_cri3}]{%
    \includegraphics[width=0.195\linewidth]{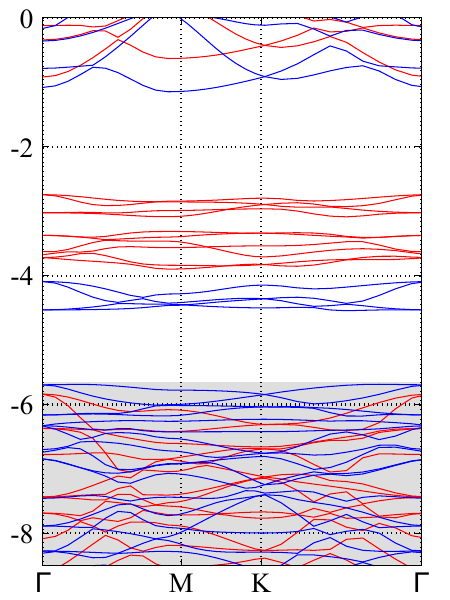}%
  }
  \hfill
  \subfloat[CrCl$_3$.\label{fig:band_align_crcl3}]{%
    \includegraphics[width=0.195\linewidth]{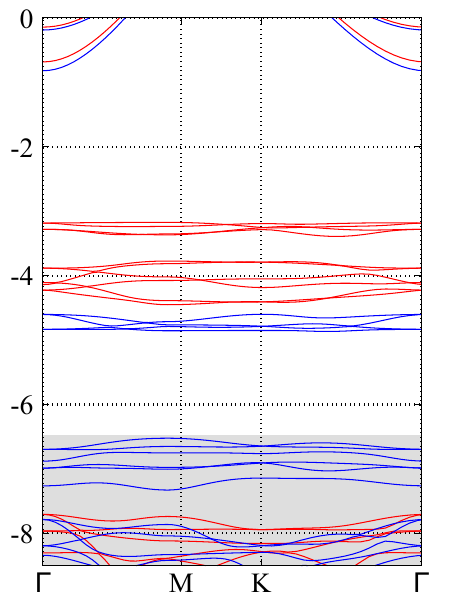}%
  }
  \subfloat[CrSBr.\label{fig:band_align_crsbr}]{%
    \includegraphics[width=0.195\linewidth]{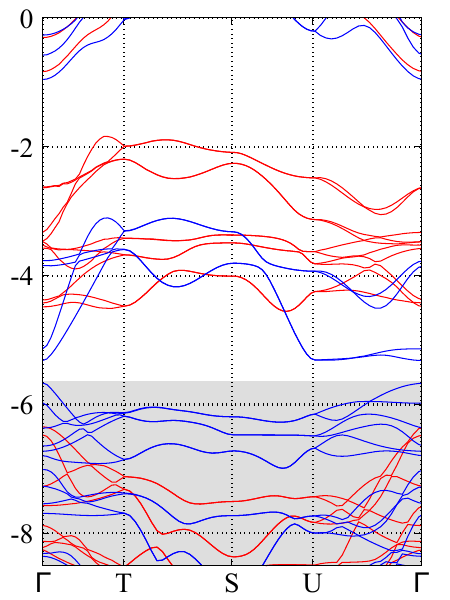}%
  }
  \hfill
  \subfloat[hBN $V_B^-$ defect.\label{fig:band_align_vb-}]{%
    \includegraphics[width=0.195\linewidth]{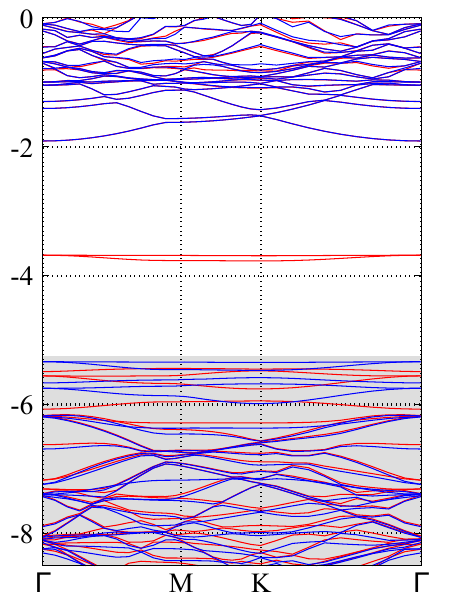}%
  }
  \caption{The PBE-calculated spin-polarized band structure, showing majority spin (blue) and 
    minority spin (red) for
    (a) hBN $V_B$, (b) CrI$_3$, 
    (c) CrCl$_3$, (d) CrSBr, and (e) hBN $V_B^-$. 
    Monolayer structures with a $16$ \AA{\ }vacuum layer are adopted. The vacuum level is set to zero for all band structures (a-d). The grey shaded regions represent the valence bands.}
  \label{fig:bandalignment}
\end{figure*}

\subsection{Band Alignment}

Since $V_B^-$ center is negatively charged, electron 
transfer between $V_B^-$ center to the sensing target can lead to change of charge state and
eventually the instability of the spin center.
To satisfy the first design principle and ensure the the stability of $V_B^-$ center, it is imperative to avoid a type-III
(or broken-gap) band alignment between the sensing target and $V_B^-$ center. 
In the type-III configuration, the highest occupied molecular orbital (HOMO) of $V_B^-$ center
overlaps with or exceeds the lowest unoccupied molecular orbital (LUMO) of the 
sensing target, or vice versa. 
This would suggest an alternative electronic ground state other than the desired $V_B^-$ state.
Fig. \ref{fig:schematic_bandalignment} shows an example of preferred band alignment between the sensing target and $V_B^-$ center.

We expect proper band alignment to serve as a baseline requirement on selecting specific candidate sensing targets.
As shown in Fig. \ref{fig:bandalignment}, band structure calculations are carried out for 
$V_B$ and candidate monolayer target magnets including CrI$_3$, $\text{CrCl}_3$, and CrSBr.
The band energies are defined relative to vacuum level. 
In the case of $V_B^-$, the vacuum level of this stand-alone charged defect 
is not well-defined due to the slow $1/r$ decay of the charge's Coulomb potential. 
However, when the $V_B^-$ and target magnets are placed adjacent to each other, the Coulomb potential will act on both systems despite a small drop.
An estimate to the band alignment may be obtained by neglecting the drop, and aligning the hBN bulk band of $V_B^-$ (Fig. \ref{fig:band_align_vb-}) to those obtained from $V_B$ (Fig. \ref{fig:band_align_vb}).
From the alignment, we find that the LUMO of CrI$_3$ (Fig. \ref{fig:band_align_cri3})
and CrCl$_3$ (Fig. \ref{fig:band_align_crcl3}) sit above the HOMO
of $V_B^-$ (Fig. \ref{fig:band_align_vb-}). 
This assures the charge state of $V_B^-$ will be stable when stacked in close proximity
to CrI$_3$ and $\text{CrCl}_3$.
In contrast, the LUMO of CrSBr (Fig. \ref{fig:band_align_crsbr}) overlaps
in energy with the HOMO of $V_B^-$ center, 
indicating charge transfer between $V_B^-$ and CrSBr.

Based on this guideline, we performed direct calculations to verify the stability of $V_B^-$ center
when placed on various 2D magnetic semiconductors that can serve as candidate sensing targets. 
As an example, we construct a commensurate heterostructure with a layer of hBN, containing one $V_B^-$ center,
and a monolayer of CrI$_3$, positioning the $V_B^-$ center directly above a Cr atom.
After full supercell lattice relaxation, the heterostructure remains flat, with no discernible out-of-plane 
displacements or bends. 
The $V_B^-$ center retained a
total magnetic moment of $2 \mu_B$, verifying the stability of desired $V_B^-$ state.
In addition, there is no significant alteration in the bonding environment and
the interlayer interaction remains predominantly vdW in nature. 

Similar test of stability is also performed for CrCl$_3$ and CrSBr. The heterostructure formed by
$V_B^-$ center and monolayer CrCl$_3$ displays the same structural and electronic stability as
shown in the case of CrI$_3$. On the other hand, the heterostructure formed by $V_B^-$ center
and monolayer CrSBr fails to maintain both the atomic and electronic structure of $V_B^-$ state
after full lattice relaxation. Our band alignment requirement successfully estimates the stability
in all three test cases.

For the following study, we selected CrI$_3$ due to the considerable band gap it offers in the $V_B^-$
and magnetic substrate heterostructure system, ensuring higher electronic stability of the $V_B^-$ charge state. 
We note that the self-energy corrections due to many-electron effects are not well captured in the 
Kohn-Sham band structures. 
To better describe the band alignment, we have further performed 
calculations using hybrid functional HSE06 \cite{heyd_hybrid_2003, krukau_influence_2006}
for its reasonable computational cost in the supercell. 
The HSE06 band alignment results (see Supporting Information) 
are consistent with DFT-PBE band alignment. 

\subsection{Effective Exchange Field}
The second design principle of proximity quantum sensing requires significant exchange interaction 
between the quantum sensor and target.
Monolayer CrI$_3$ displays an out-of-plane easy-axis ferromagnetic ground state under strong intrinsic spin-orbit coupling \cite{xu_interplay_2018}. 
Given negligible spin-orbit coupling strength in hBN, the spin orientation of $V_B^-$ spin
center should be polarized and collinear with the out-of-plane easy axis of CrI$_3$.
The exchange interaction between CrI$_3$ and $V_B^-$ can be described by a spin Hamiltonian with an effective exchange field $\textbf{B}_{\text{exc}}$,
\begin{equation}
    H = g \mu_B \textbf{B}_{\text{exc}} \cdot \textbf{S}
\end{equation}
where $\textbf{S}$ is the total electron spin-1 operator for $V_B^-$ center,
$g$ is the Landé factor, $\mu_B$ is the Bohr magneton.
In our collinear calculation, $\textbf{B}_{\text{exc}}$ remains aligned to the direction
of the triplet spin center. Thus, the magnitude of such effective exchange field can be obtained by,
\begin{equation}
    |\textbf{B}_{\text{exc}}| = \frac{1}{g \mu_B \Delta m_s}(\text{E}_\text{FM} - \text{E}_\text{AFM})
\end{equation}
where E$_\text{FM}$ (E$_\text{AFM}$) is the total energy of the hybrid system with
FM (AFM) coupling between $V_B^-$ spin center
and CrI$_3$ substrate. $\Delta m_s$ is the change of spin quantum number of $V_B^-$.

While the sign and magnitude of this effective exchange field can be postulated to vary, 
contingent on the stacking configuration between the hBN layer and CrI$_3$ substrate,
our first-principles calculations suggest that 
the interaction always favors AFM coupling across all stacking configurations, regardless of interlayer twist angle and lateral shift.
To demonstrate this, we adopt a hBN flake structure hosting a $V_B^-$ defect at the center with 
hydrogen termination on the edge to model the quantum sensor. 
The flake structure (in comparison to extended structure) allows 
us to fully investigate the rotational and translational
degree of freedom of the stacking dependence, free from the geometry constraint of supercells.

\begin{figure}[h]
  \centering
  \subfloat[\label{fig:heatmap_schematic}]{%
    \includegraphics[width=0.5\linewidth]{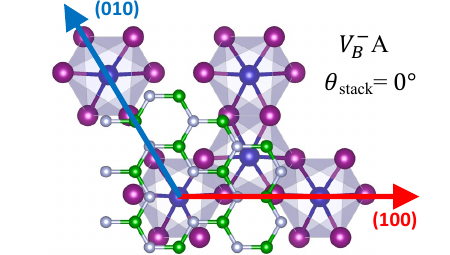}%
  }
  \hfill
  \subfloat[\label{fig:1st_flake_angle_0_heatmap}]{%
    \includegraphics[width=0.5\linewidth]{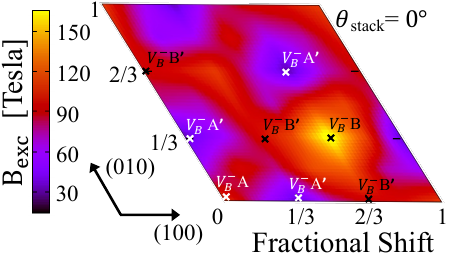}%
  }
  \hfill
  \subfloat[\label{fig:1st_flake_angle_90_heatmap}]{%
    \includegraphics[width=0.5\linewidth]{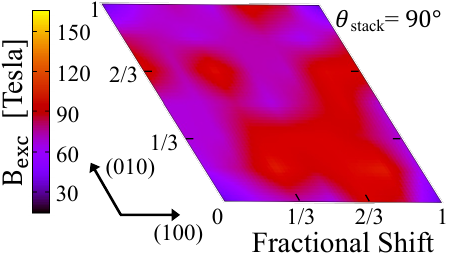}%
  }
  \hfill
  \subfloat[\label{fig:1st_layer_hetero}]{%
    \includegraphics[width=0.5\linewidth]{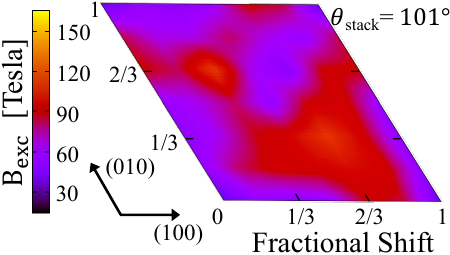}%
  }
  \caption{Dependence of effective exchange field $\textbf{B}_{\text{exc}}$ on interlayer stacking configuration of monolayer hBN/CrI$_3$. 
  (a) Top view of a $V_B^-$ defect positioned directly above a Cr atom in the A sublattice of the
  CrI$_3$ unit cell ($V_B^-$A stacking) which establishes the reference configuration for both the interlayer rotation $\theta_{\text{stack}}$ and lateral translation. The (100) and (010) lateral shift directions are labeled with red and blue arrows, respectively. 
  Effective exchange field $\textbf{B}_{\text{exc}}$ calculated using hBN flake method (b) $\theta_{\text{stack}} = 0^\circ$ 
  and (c) $\theta_{\text{stack}} = 90^\circ$. (d) $\textbf{B}_{\text{exc}}$ calculated
  using a periodic heterostructure 
  with $\theta_{\text{stack}} = 101^\circ$. 
  $V_B^-$B, $V_B^-$B', $V_B^-$A' stacking order corresponds to a fractional lateral
  shift of hBN flake by approximately (2/3, 1/3), (2/3, 0), (1/3, 0) with respect to the $V_B^-$A stacking, respectively. 
  The heatmaps in (b-d) were drawn by interpolating neighboring data points on a 6x6 grid. }
  \label{fig:1st_layer_heatmap}
\end{figure}

We start by considering a monolayer hBN with $V_B^-$ in direct contact with CrI$_3$ (shown in 
Fig. \ref{fig:heatmap_schematic}). 
As expected from our previous discussions, $V_B^-$ defect remains structurally and 
electronically stable upon stacked on CrI$_3$.
Due to the $C_3$ symmetry of $V_B^-$, the heterostructure's full interlayer twisting 
degree of freedom can be reduced to the range of $\theta_{\text{stack}} \in [0, \frac{2 \pi}{3}]$,
where $\theta_{\text{stack}}$ is the relative in-plane rotation angles between hBN and CrI$_3$.
In total, 4 unique twist angles $\theta_{\text{stack}}$, each with a $6 \times 6$ grid for 
lateral translation in the
unit cell of CrI$_3$, were sampled.
The pattern of effective exchange field at $\theta_{\text{stack}} = 0^\circ$ and $90^\circ$ are shown 
in Fig. \ref{fig:1st_flake_angle_0_heatmap} and Fig. \ref{fig:1st_flake_angle_90_heatmap}, respectively. 
(see Supporting Information for data of $\theta_{\text{stack}} = 0^\circ, 30^\circ,
60^\circ, 90^\circ$).

In order to verify that the findings obtained with the hBN flake method are not influenced by 
finite size effects, we also performed similar calculations using a periodic heterostructure 
comprised of hBN and CrI$_3$. 
A $6 \times 6$ grid within the unit cell of CrI$_3$ is sampled and the heatmap of its effective
exchange field is shown in Fig. \ref{fig:1st_layer_hetero}. 
The magnitude and direction of $\textbf{B}_{\text{exc}}$ calculated using periodic heterostructure at $101^\circ$
is close to those obtained from the hBN flake method at $90^\circ$.

The average of effective exchange field over the sampled stacking configurations is
$\overline{\textbf{B}}_{\text{exc}} \approx 83$ T (equivalently $\text{E}_\text{FM} - \text{E}_\text{AFM} \approx 10$ meV), 
several orders of magnitude greater than typical size of magnetic dipole interaction.
This value varies with a standard deviation of $\sigma_{std} = 19$ T, ranging between
a minimum of $33$ T and a maximum of $168$ T.
Interestingly, the direction of the exchange field, or the sign of $\textbf{B}_{\text{exc}}$ from the calculations, 
stays unchanged throughout all stacking configurations. 
This contrasts with the spatially varying direction of stray fields generated by a 
2D magnet at several angstroms away from the sample surface, as we will discuss later.
Typically, quantum sensing protocol conducted with solid-state spin centers, 
such as the $NV^-$ center in diamond, is performed by measuring the projected stray field to the spin 
center's quantization axis.
Spatial variation in direction can cause the projection, and hence the interaction magnitude, 
to fluctuate around zero, effectively hindering sensing at atomic level resolution.
Relying on a fundamentally different mechanism, our proposed protocol showcases unidirectional AFM exchange field that promises
high sensitivity in probing the magnetism.

\subsection{Exchange Pathway}

Next we discuss the dependence of the magnitude of  
$\textbf{B}_{\text{exc}}$ on stacking order.
We selectively highlighted two equivalent stacking orders, $V_B^-$A' and $V_B^-$B', 
in Fig. \ref{fig:1st_flake_angle_0_heatmap}. These particular stacking configurations were
chosen for illustration because they respectively exhibits comparable magnitudes of 
$\textbf{B}_{\text{exc}}$ in all the $\theta_{\text{stack}}$ we calculated, suggesting
a strong correlation between stacking order and $\textbf{B}_{\text{exc}}$. 

\begin{figure}[h]
  \centering
  \subfloat[\label{fig:excPath_vesta}]{%
    \includegraphics[width=0.4\linewidth]{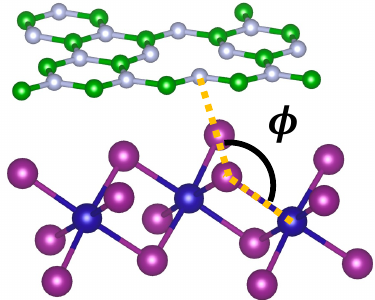}%
  }
  \hfill
  \subfloat[\label{fig:excPath_orbital}]{%
    \includegraphics[width=0.55\linewidth]{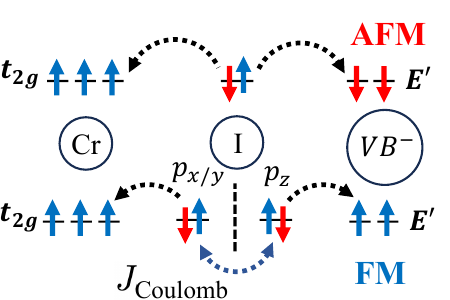}%
  }
  \caption{(a) Schematic of interlayer superexchange pathway (orange dashed line) between the Cr atom and $V_B^-$ defect with the exchange pathway angle $\phi$ shown. (b) Schematic of the angle dependent AFM/FM interlayer superexchange.}
  \label{fig:excPath}
\end{figure}

The microscopic mechanism of the interlayer coupling is an AFM superexchange between the Cr atoms
in CrI$_3$ and $V_B^-$ defect mediated by the I p-orbitals.
The Cr atom in CrI$_3$ is in a $3d^3$ electronic configuration with 3 unpaired electrons in the $t_{2g}$ 
orbitals at the valence top. 
The $V_B^-$ defect's magnetic moment originates from 2 half filled $E'$ orbitals \cite{abdi_color_2018}.
Many exchange pathways between the defect and Cr are possible.
We select six main superexchange pathways, which include nearest-neighbor 
I atoms in coordination with N atoms surrounding the $V_B^-$ defect and two adjacent Cr 
atoms, with an angle $\phi$ in each pathway, one of such pathway is illustrated in Fig. \ref{fig:excPath_vesta}.
Fig. \ref{fig:excPath_orbital} demonstrates competing exchange interaction process between Cr atom and $V_B^-$ defect.
When the angle $\phi$ of the exchange pathway 
approaches $180^{\circ}$, both Cr $t_{2g}$ orbitals and the
$V_B^-$ $E'$ orbitals couple to the same I p-orbital, which leads to an AFM superexchange. 
Conversely, as $\phi$ nears $90^{\circ}$, the Cr $t_{2g}$ orbitals and 
$V_B^-$ $E'$ orbitals tend to couple to orthogonal I p-orbitals, favoring
an interlayer FM superexchange via Coulomb exchange interactions on the I atom. 
This competition between interlayer AFM and FM superexchange is known as the Goodenough-Kanamori rule 
\cite{goodenough_theory_1955, goodenough_interpretation_1958, kanamori_superexchange_1959}.
We then calculated the mean value of $\phi$ under different stacking configurations and found 
$\phi$ and $\textbf{B}_{\text{exc}}$ are positively correlated,
with a notable coefficient of determination of $R^2 = 0.5$
(see Supporting Information for detail). 
This strong correlation substantiates our theoretical analysis on the nature of
interlayer superexchange interaction, which favors stronger AFM coupling as $\phi$ increases.
We do no find including Hubbard U term to have pronounced influence on the property of this 
proximity-induced superexchange mechanism (see Supporting Information for more detail).

\subsection{Sensing Across Multilayer hBN}

Given the significant exchange field when the $V_B^-$ center is situated 
in the first hBN layer above the magnetic substrate, 
we investigate a realistic scenario in which the defect spin center is separated by pristine BN layers from the sensing targets.
This scenario mimics experiments where multiple layers BN are often stacked atop vdW magnets.
We sample stacking configurations for a bilayer hBN (in its natural AA' stacking)
flake on the CrI$_3$ heterostructure similar to the case of monolayer hBN/CrI$_3$, but with an additional pristine hBN layer 
between the $V_B^-$ defect layer and the CrI$_3$ substrate. 

\begin{figure}[h]
  \subfloat[\label{fig:2nd_flake_angle_0_heatmap}]{%
    \includegraphics[width=0.5\linewidth]{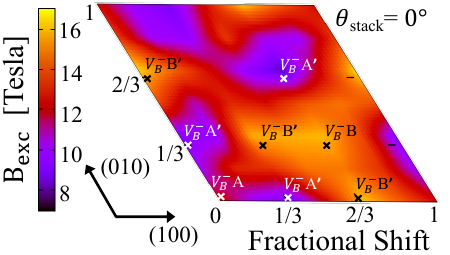}%
  }
  \hfill
  \subfloat[\label{fig:2nd_flake_angle_90_heatmap}]{%
    \includegraphics[width=0.5\linewidth]{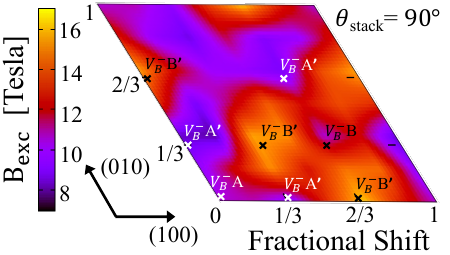}%
  }
  \caption{Effective exchange field $\textbf{B}_{\text{exc}}$ calculated 
  for the bilayer-hBN CrI$_3$ system, where there is a pristine hBN between $V_B^-$ layer 
  and CrI$_3$ substrate. Full space lateral translation under certain $\theta_{\text{stack}}$
  interlayer relative rotation.
  (a) $\theta_{\text{stack}} = 0^\circ$ 
  and (b) $\theta_{\text{stack}} = 90^\circ$. Since hBN has a AA' ground state stacking, various
  stacking orders are preserved as marked in the heatmap.}
  \label{fig:2nd_layer_heatmap}
\end{figure}

Despite the much larger spatial separation between the $V_B^-$ and CrI$_3$, we discover large exchange 
field that always favors AFM coupling in the bilayer hBN/CrI$_3$ system. 
The field averages to $\overline{\textbf{B}}_{\text{exc}} \approx 12$ T
(equivalently $\text{E}_\text{FM} - \text{E}_\text{AFM} \approx 1.4$ meV), with $\sigma_{std} = 2.3$ T and ranging between $8$ T and $16$ T.
This is a surprising finding because the exchange pathway involves an additional 
pristine layer of hBN with large band gap, 
which introduces a barrier for the virtual hopping across the vdW gap. 
Fig. \ref{fig:2nd_layer_heatmap} shows the 
magnitude of $\textbf{B}_{\text{exc}}$ when the $V_B^-$ defect is stacked onto 
CrI$_3$ at different position. (see Supporting Information for full data)
The exchange field maintains a similar magnitude
across varied interlayer rotations and translations, exhibiting patterns consistent 
with those we discovered from first layer. 

To understand the exchange interactions, we notice that with an extra pristine hBN layer, interlayer Cr-$V_B^-$ exchange is mediated
by the I p-orbital and the pristine hBN's p-orbital, therefore super-super-exchange in nature.
Given the natural stacking configuration of the hBN layers, the vertical interlayer hopping between the orbitals of the defect and those in the pristine BN layers remains unchanged 
despite shifts and rotations of the bilayer relative to the CrI$_3$ substrate. 
It is the hybridization between the pristine hBN and the CrI$_3$ that determines
the variation of orbital-dependent FM and AFM exchange competition. 
In comparison to the superexchange process, the super-super-exchange pathway now involves a B atom 
directly beneath a N atom adjacent to the $V_B^-$, extending to an I atom, and finally connecting
to the nearest neighbor Cr atom. 
Our first-principle results suggest that the characteristic AFM coupling, facilitated
by the higher-order super-exchange 
through the layered hBN matrix, should be preserved even as the thickness of pristine hBN between the 
$V_B^-$ defect layer and CrI$_3$ substrate increases, since it only contributes additional vertical B-N hoppings with no stacking dependence.
Therefore, our proposed quantum sensing protocol driven by such exchange interaction can be
extrapolated to structures incorporating more than one layer of pristine hBN, although systematic first-principles studies are unfeasible at this moment due to the higher computational cost. 

\subsection{Comparison to Stray Field Sensing}

\begin{figure}
    \includegraphics[width=0.95\linewidth]{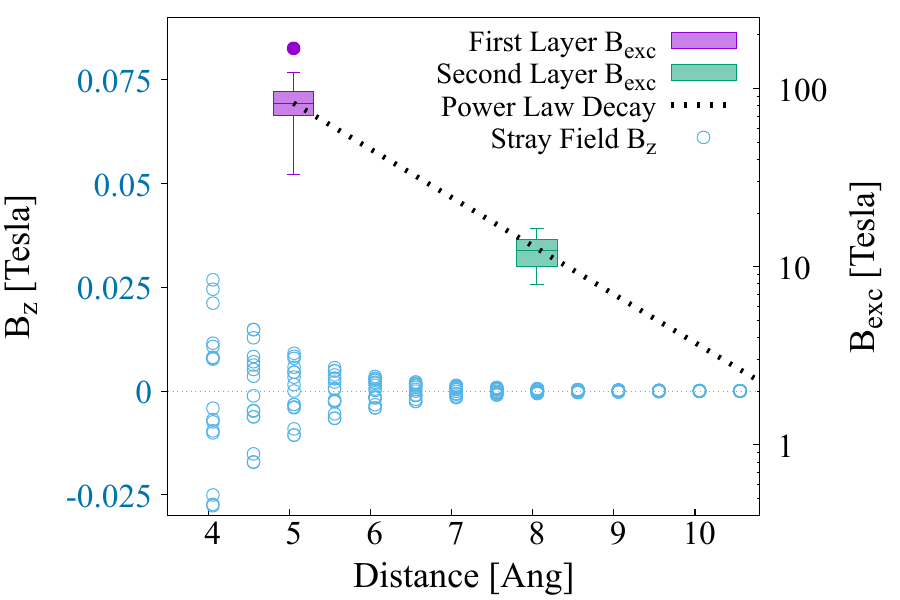}
    \caption{\label{fig:layer_dependence} Quartile box plot for the magnitude of effective exchange 
    field $\textbf{B}_{\text{exc}}$ on log scale containing 
    all sampled stacking configurations from $V_B^-$/CrI$_3$ heterostructure.
    Dashed black line extrapolates the power law decay of exchange interaction strength.
    Stray field's $\hat{z}$ component, $B_z$, is shown in linear scale (blue) calculated
    from uniformly sampled positions within a CrI$_3$ unit cell.}
\end{figure}

Furthermore, we analyze the dominance of exchange interaction over dipole-dipole interaction 
between the $V_B^-$ spin center and the magnetic substrate as the proximity changes. 
Fig. \ref{fig:layer_dependence} represents 
the exchange interaction $\textbf{B}_{\text{exc}}$ of various stacking configurations
within the first two hBN layers and compares it with the $\hat{z}$ component of 
stray field $B_z$ generated by CrI$_3$ substrate that is periodic in the basal plane,
\begin{equation}
    \mathbf{B}(\mathbf{r}) = \frac{\mu_0}{4 \pi} \biggr[ \frac{3\mathbf{r}(\mathbf{m}\cdot \mathbf{r})}{r^5} -\frac{\mathbf{m}}{r^3}\biggr]
\end{equation}
where $\mu_0$ is the vacuum permeability and $\mathbf{m}$ is the magnetic moment of Cr atom.
The stray field, arising from each Cr ion's $3 \mu_B$ out-of-plane magnetic moment, is 
three orders of magnitude weaker than $\textbf{B}_{\text{exc}}$ at close proximity. 
Moreover, while $\textbf{B}_{\text{exc}}$ remains robustly AFM, $B_z$ exhibits
spatial variations in sign, evident from its fluctuations around zero in Fig. \ref{fig:layer_dependence}.

The robust magnetic proximity effect in the presence of additional hBN layers suggests that 
the exchange interaction can dominate over dipole field even at larger distance. 
As the exchange interaction depends on the overlap of electronic wavefunction or hopping process, its magnitude
should exhibit a power law decay as the distance or pathways become longer. 
Based on the decay rate from first to second layer of hBN,
we anticipate that $\textbf{B}_{\text{exc}}$ will reach the order of 0.1 mT at the seventh layer, 
thereby still maintaining its dominance over the stray field for spin center embedded in 3D structure. 

\section{Summary}

In this study, we introduces a novel quantum sensing scheme driven by the exchange interaction induced 
by the proximity between the spin center and magnetic sensing target. The practical implementation of 
this principle was explored utilizing the $V_B^-$ spin center in hBN proximate to a CrI$_3$ magnetic 
substrate, employing \textit{ab-initio} methods. 
Our calculation revealed a unidirectional and robust AFM exchange interaction between $V_B^-$ and CrI$_3$.
The exchange field is estimated to dominate over the magnetic dipole field
with up to seven layers of pristine hBN between the spin center and the magnet. 

As controlled generation of $V_B^-$ defect at different depths and positions has been experimentally 
demonstrated \cite{guo_generation_2022, gao_high-contrast_2021},
our proposed proximity-induced exchange interaction driven quantum sensing protocol can be readily realized 
using hBN coated cantilever or in fabricated heterostructure device. 
The principles established from the $V_B^-$ and CrI$_3$ case may be applied to other 
optically active spin defects in 2D materials, such as WS$_2$ \cite{li_carbon_2022}. 
On the other hand, the $V_B^-$ defect may also be applied to investigate magnetic materials with 
zero net magnetization, such as AFM materials. 
This is because the exchange interaction is primarily dominated by the direct or indirect overlap of electronic wavefunctions between quantum sensors and sensing targets, 
and the $V_B^-$ defect strongly couples to magnetic ions or layers in its immediate proximity, rather than the total dipole field.
This study thereby pioneers a ultrasensitive \textit{in situ} quantum sensing model driven
by proximity-induced exchange interaction between quantum sensor and target, which may overcome intrinsic limitations of stray field sensing.

\section{Methods}
All \textit{ab-initio} calculations (except those specifically noted)
were performed by using Density Functional Theory (DFT), with the 
Perdew-Burke-Ernzerhof (PBE) \cite{perdew_generalized_1996}
functional for electron exchange and correlation 
potentials, as implemented in the VASP code \cite{kresse_efficiency_1996}. 
We employed the projector augmented wave (PAW) \cite{kresse_ultrasoft_1999}
method for electron-ion interaction and an energy cutoff of 520 eV for wavefunction. Atomic
coordinates are relaxed until forces on atom are smaller than $0.01$ eV/\AA{\ } and the total energy
was converged to be within $10^{-6}$ eV. Van der Waal interactions is included as the DFT-D2 method
\cite{grimme_semiempirical_2006}.
Moreover, a $16$ \AA{\ }vacuum layer is adopted to avoid interactions between repeating images. 

For band alignment calculation, a $5 \times 5$ hBN supercell containing one boron vacancy
was used, with hBN lattice constant $2.51$ \AA{\ } and
$\mathbf{\Gamma}$ centered $6 \times 6 \times 1$ k-point sampling.
Monolayer unit cell of CrI$_3$ and CrCl$_3$ was used with lattice constant $7.01$ \AA{\ } and 
$5.97$ \AA{\ } respectively, with a $\mathbf{\Gamma}$ centered $9 \times 9 \times 1$ k-point sampling
for both. We use monolayer unit cell of CrSBr with lattice constant
$a = 3.51$ \AA{\ } and $b = 4.71$ \AA{\ } with
$\mathbf{\Gamma}$ centered $12 \times 9 \times 1$ k-point sampling.

For the effective exchange field calculation, we use a circular flake structure with 41 B and
42 N atoms, containing one boron vacancy in the center
and hydrogen termination on the edge. The underlying substrate contains a $4 \times 4$ CrI$_3$
unit cell in each periodic unit. We performed $\mathbf{\Gamma}$ only k-point calculation.
The alternative periodic supercell of hBN and CrI$_3$ heterostructure is formed by
a layer of 16 hBN unit cell on 4 CrI$_3$ unit cell with $\mathbf{\Gamma}$ centered 
$3 \times 3 \times 1$ k-point sampling.

\section{Supporting Information}
HSE06 band alignment, formation energy heatmap, correlation between exchange pathway angle $\phi$ and effective exchange field $\textbf{B}_{\text{exc}}$,
influence of correlation effect
on exchange interaction, POSCAR for VASP input, and raw data for effective exchange field.

\section{Acknowledgments}
We thank Zeeshawn Kazi, Christian Pederson, Vasileios Niaouris, and Kai-Mei Fu for
insightful discussions.
The theoretical framework of quantum sensing at proximity is supported by the U.S. Department of Energy, Office of Science, National Quantum
Information Science Research Centers, Co-design Center for 
Quantum Advantage (C2QA) under contract number DE-SC0012704. 
The first-principles investigation of magnetism and exchange coupling is based upon work supported by the National Science Foundation under Award No. DMR-2339995.
This work was facilitated through the use of advanced computational, storage, and networking infrastructure provided by the Hyak supercomputer system and funded by the University of Washington Molecular Engineering Materials Center at the University of Washington. (NSF MRSEC DMR-2308979). 
This material is based in part upon work supported by the state of Washington through the University of Washington Clean Energy Institute.

\bibliography{apssamp}
\bibliographystyle{achemso}

\end{document}


\title{Supporting Information: Proximity-Induced Exchange Interaction: a New Pathway for Quantum Sensing using Spin Centers in Hexagonal Boron Nitride}

\author{Lingnan Shen}
    \affiliation{Department of Physics, University of Washington, Seattle, WA, USA}
\author{Di Xiao}
    \email{dixiao@uw.edu}
    \affiliation{Department of Material Science and Engineering, University of Washington, Seattle, WA, USA}
    \affiliation{Department of Physics, University of Washington, Seattle, WA, USA}
    \affiliation{Pacific Northwest National Laboratory, Richland, WA, USA}
\author{Ting Cao}
    \email{tingcao@uw.edu}
    \affiliation{Department of Material Science and Engineering, University of Washington, Seattle, WA, USA}

\date{\today}

\maketitle

\onecolumngrid
\tableofcontents

\clearpage

\section{HSE06 Band Alignment}
\begin{figure*}[h!]
  \centering
  \subfloat[hBN $V_B$ defect]{%
    \includegraphics[width=0.218\linewidth]{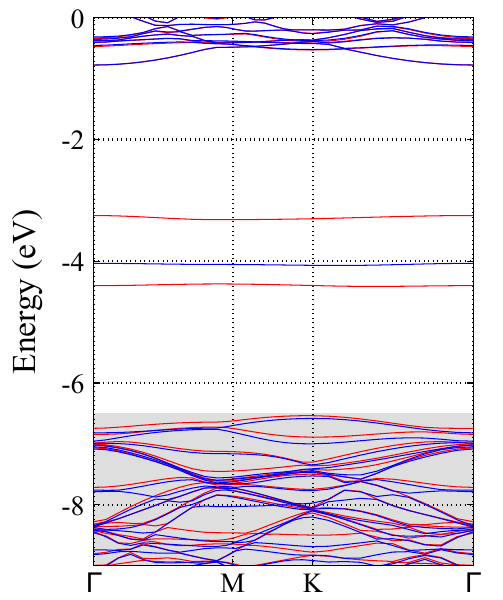}%
  }
  \hfill
  \subfloat[CrI$_3$]{%
    \includegraphics[width=0.195\linewidth]{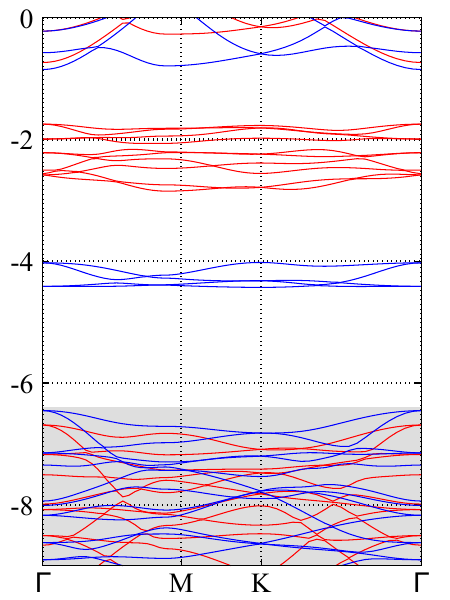}%
  }
  \hfill
  \subfloat[CrCl$_3$]{%
    \includegraphics[width=0.195\linewidth]{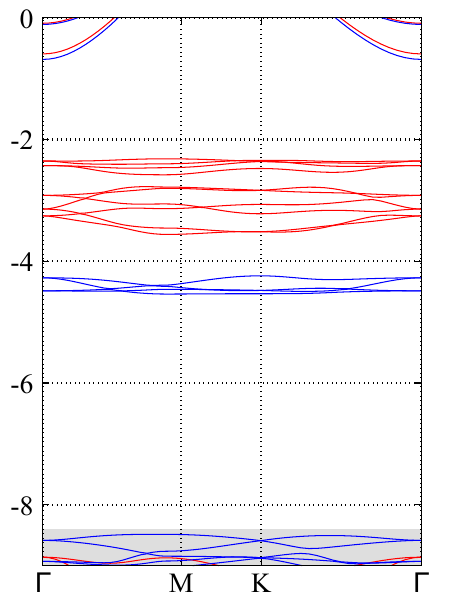}%
  }
  \subfloat[CrSBr]{%
    \includegraphics[width=0.195\linewidth]{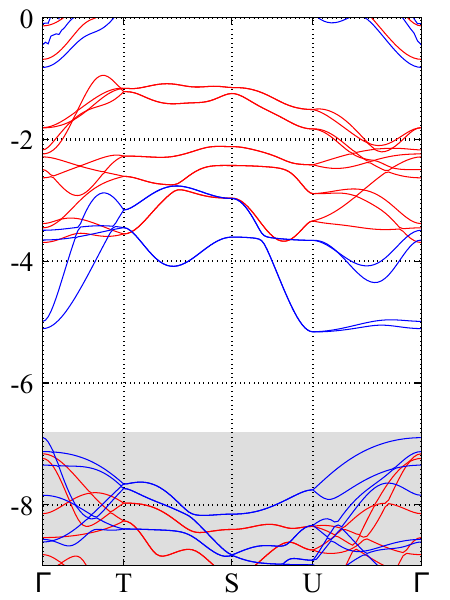}%
  }
  \hfill
  \subfloat[hBN $V_B^-$ defect]{%
    \includegraphics[width=0.195\linewidth]{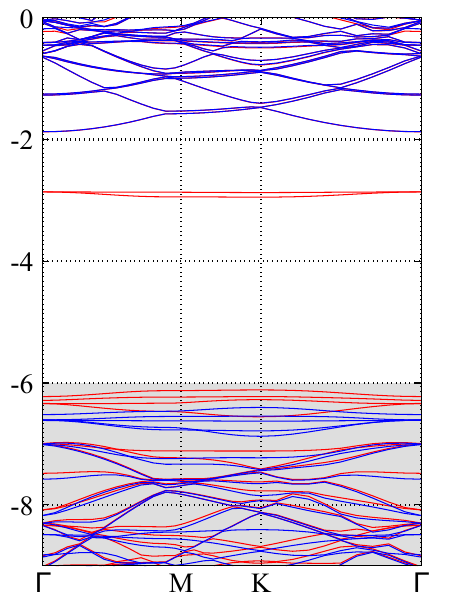}%
  }
  \caption{The HSE06-calculated spin-polarized band structure, showing majority spin (blue) 
    and minority spin (red)
    for (a) hBN $V_B$, (b) CrI$_3$, 
    (c) CrCl$_3$, (d) CrSBr, and (e) hBN $V_B^-$. 
    Monolayer structures with a $16$ \AA{\ }vacuum layer are adopted and the vacuum level is set to zero for all band structures (a-e). The grey shaded regions represent the valence bands.}
  \label{fig:bandalignment}
\end{figure*}


\section{Exchange Pathway Angle $\phi$ and Effective Exchange Field $\textbf{B}_{\text{exc}}$}

To investigate the correlation between stacking configuration and the effective exchange field, 
we consider the nearest-neighbor 
I atom in coordination with three N atoms surrounding the $V_B^-$ defect and two adjacent Cr 
atoms, defining an angle $\phi$. We performed multilinear regression between $\phi$ and 
$\textbf{B}_{\text{exc}}$ and obtained a coefficient of determination
$R^2 = 0.5$, encompassing all sampled stacking configurations within our investigation.

\begin{figure}[h!]
    \includegraphics[width=0.45\linewidth]{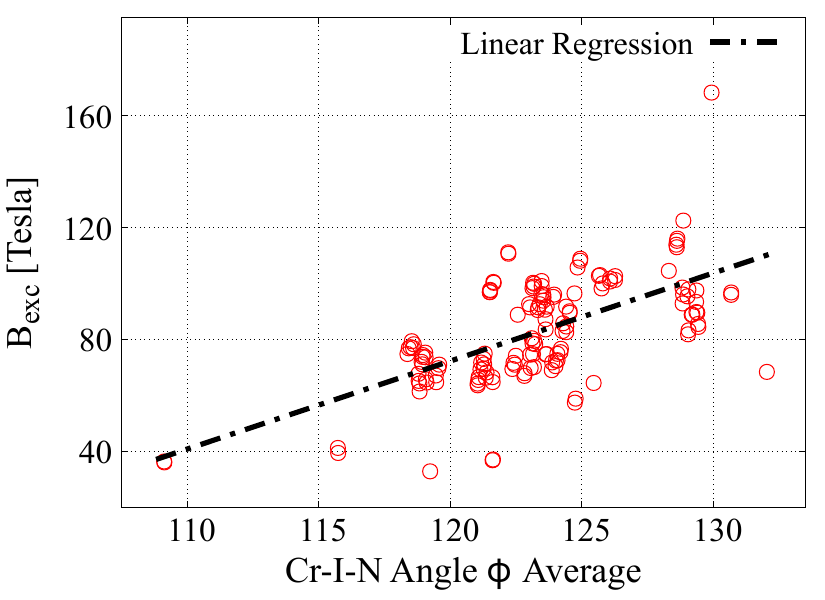}
    \caption{Linear regression analysis demonstrating the relationship between the average Cr-I-N angle, 
    $\overline{\phi}$, and the effective exchange field $\textbf{B}_{\text{exc}}$ in hBN-CrI$_3$
    heterostructure. Data points represent the $\textbf{B}_{\text{exc}}$ values corresponding to
    different stacking configurations sampled in the study, 
    with the dashed line indicating the fitted linear model.
    The regression suggests a clear trend, affirming the angular dependence of the 
    superexchange interaction strength.}
    \label{fig:excangle_correlation}
\end{figure}

In order to visualize this relationship, we plot the average of exchange pathway angle $\overline{\phi}$ 
against $\textbf{B}_{\text{exc}}$ in 
Fig. \ref{fig:excangle_correlation}. We obtained a correlation coefficient of $R^2=0.4$ with single variable 
linear regression between $\overline{\phi}$ and $\textbf{B}_{\text{exc}}$.

\clearpage

\section{Formation Energy Heatmap}
We calculated interface formation energy at $\theta_{\text{stack}} = 0^\circ$ and $60^\circ$, 
normalized by the number of BN unit cells at the interface. We did not find strong correlation between
the spatial pattern of the interface formation energy and that of the effective exchange field. 
We believe one reason for the lack of correlation is that the variation of interface formation energy 
with stacking configurations is as small as a couple meV, due to the mismatch of the two lattice constants and structures. 
This small variation does not significantly change the interlayer distance and has little effect on the exchange pathway.

\begin{figure}[h]
  \centering
  \subfloat[\label{fig:1st_flake_formationEnergy_angle_0}]{%
    \includegraphics[width=0.25\linewidth]{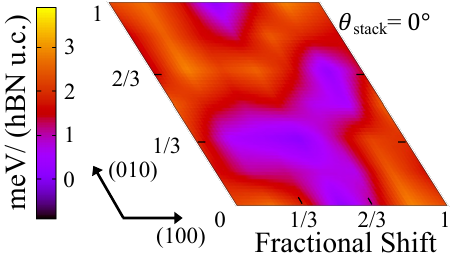}%
  }
  \hspace{1cm}
  \subfloat[\label{fig:1st_flake_formationEnergy_angle_60}]{%
    \includegraphics[width=0.25\linewidth]{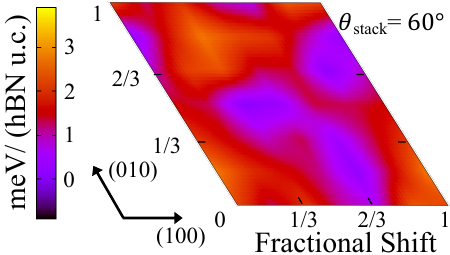}%
  }
  \caption{Dependence of interface formation energy on interlayer stacking configuration of monolayer hBN/CrI$_3$
  using hBN flake method. The zero energy is chosen as the minimum of total energy in the same $\theta_{\text{stack}}$.
  (a) $\theta_{\text{stack}} = 0^\circ$ and (b) $\theta_{\text{stack}} = 60^\circ$. 
  The heatmaps in (a-b) were drawn by interpolating neighboring data points on a 6x6 grid. }
  \label{fig:1st_layer_heatmap_formation_energy}
\end{figure}

\section{Influence of Correlation Effect}

To address the influence of correlation effect in CrI$_3$, we conducted DFT+U calculations 
for six randomly selected stacking
configurations with the monolayer hBN flake and CrI$_3$ system, applying a Hubbard $U =1.5$ eV and an 
exchange $J = 0.5$ eV for Cr atom. \cite{wu_physical_2019}
Our findings revealed a slight decrease in the average magnitude of the effective
exchange field ($\textbf{B}_{\text{exc}}$) by approximately 7.6\%, with the directionality of the 
antiferromagnetic coupling remaining consistent.
Such reduction in the exchange splitting is marginal upon incorporating the on-site 
Hubbard U correction for the Cr atom's d orbitals. 

The correlation effect in the hBN $V_B^-$ defect plays an essential role in the spin selective 
non-radiative decay 
process. However, the exchange interaction discussed in this work focuses on the ground state S=1 triplet 
manifold which is correctly described in DFT. We thus do not expect correlation effect to have pronounced 
influence on the property of this proximity induced exchange interaction.

\begin{table}[h]
    \centering
    \begin{tabular}{|c|c|c|}
        \hline
         Stacking Configuration & $\textbf{B}_{\text{exc}}$ [Tesla] (from DFT+U) & $\textbf{B}_{\text{exc}}$ [Tesla] (without U)\\
         \hline
         (0, 1/2) $\theta_{\text{stack}} = 0^\circ$ & 69.06 & 75.29\\
         \hline
         (0, 0) $\theta_{\text{stack}} = 30^\circ$ & 93.26 & 101.76\\
         \hline
         (1/2, 1/6) $\theta_{\text{stack}} = 30^\circ$ & 57.29 & 61.41\\
         \hline
         (1/3, 1/6) $\theta_{\text{stack}} = 90^\circ$ & 65.06 & 69.79\\
         \hline
         (1/3, 1/2) $\theta_{\text{stack}} = 60^\circ$ & 82.06 & 89.57\\
         \hline
         (1/3, 0) $\theta_{\text{stack}} = 0^\circ$ & 34.24 & 36.83\\
         \hline
    \end{tabular}
    \caption{Comparison of exchange field $\textbf{B}_{\text{exc}}$ calculated with Hubbard U added to the Cr atom and without.
    The configuration (x,y) represent fractional lateral shifts x along the (100) and 
    y along the (010) directions of CrI$_3$ unit cell respectively.}
    \label{tab:my_label}
\end{table}

\clearpage

\section{POSCAR for VASP Input}
Here we include the POSCAR inputs for VASP calculation for the hBN flake with CrI$_3$ method 
and commensurate hBN/CrI$_3$ heterostructure method used in our study.

The commensurate supercell calculation has an isotropic strain of 0.2\% on the hBN layer. 
We believe this artificial strain is negligible and the fundamental physics governing the circular flake and 
commensurate supercell remains the same. Our calculations of exchange field in the two cases attest to this conclusion. 

The hBN flake containing one $V_B^-$ defect and CrI$_3$ input is:
\begin{verbatim}
vb_q-1_hbn_flake_crI3
1.00000000000000
    28.0399999999999991    0.0000000000000000    0.0000000000000000
     0.0000000000000000   24.2833499999999987    0.0000000000000000
     0.0000000000000000    0.0000000000000000   23.0000000000000000
B    N    H    Cr   I 
41    42    24    32    96
Direct
0.58151825377342 0.24065008854338 0.588688513201531
0.65882670202234 0.29225930520822 0.590160922914537
0.42476377595971 0.24093424074434 0.588700817397322
0.50341825178416 0.29237200379058 0.588626933215444
0.58107293900650 0.34451662760490 0.588783361694163
0.65892692089667 0.39621336587418 0.589823923274211
0.73677829197519 0.44927513459900 0.590539306261202
0.34744653533633 0.29301948312003 0.589964224307276
0.42541680556931 0.34455188092386 0.588265311871364
0.50303182290205 0.39702741588641 0.588547294773514
0.58058732848751 0.44878494414637 0.589100985102632
0.65868266166636 0.50027894823227 0.589255910694907
0.73622555450300 0.55259442779350 0.591175022199855
0.26973935541048 0.34422755424312 0.591098747704111
0.34718066338517 0.39657036487547 0.588846023925147
0.42574118573048 0.44844960768006 0.588535412224081
0.58056115169475 0.55177956836405 0.588859246935843
0.65869764964507 0.60452019962002 0.589481469910037
0.73660262571885 0.65592894028819 0.591764340608527
0.26922797146324 0.44845130878919 0.588735095932914
0.34736851805126 0.50020440807158 0.588669867823665
0.42578100837948 0.55186228462661 0.588606941330521
0.50295475013491 0.60340627347324 0.588847222744143
0.58084549175351 0.65604232961141 0.588705379118076
0.65867367201402 0.70840159975182 0.588576383330323
0.19139420303269 0.50071958617623 0.588027265146211
0.26944471658451 0.55208673971183 0.589903192078157
0.34731450432163 0.60385478507264 0.589777524384760
0.42526143270356 0.65576353221999 0.589058012375601
0.50314117885905 0.70784429480884 0.589368541662386
0.58086016421683 0.75992415859064 0.589273018193922
0.65836936381736 0.81238344132860 0.587234793116507
0.19148363720070 0.60364826477995 0.592634802937383
0.26931230966228 0.65569049694585 0.592857370905892
0.34728247114226 0.70751653812107 0.590822622071174
0.42537497742556 0.75963237980807 0.590531810207953
0.50321798240582 0.81175662692413 0.591860661708701
0.58113823326481 0.86366051799991 0.591157663076906
0.26994367482815 0.75986710266790 0.591739401997047
0.34754241124405 0.81105346979993 0.590677049118603
0.42477426819857 0.86315003649021 0.591258578858667
0.53084234414923 0.24198570036844 0.588598744372099
0.60726941912444 0.29218866249247 0.589107922889615
0.68309150385176 0.34365170450691 0.590727450218115
0.37394684314301 0.24223970290490 0.589373466657364
0.45114344133919 0.29187902765158 0.588295197740127
0.52921675250394 0.34399135733599 0.588533637548176
0.60735026577067 0.39612566845787 0.588809910270260
0.68536011594712 0.44827118219535 0.590030149192577
0.76116306868082 0.50077544891715 0.590984880626725
0.29605758733055 0.29416829346166 0.591450557975012
0.37323699555792 0.34462203574955 0.588921510827505
0.45097630626818 0.39585044662662 0.587909184831816
0.53065060005154 0.44517136377599 0.589109082599060
0.60743835915344 0.50029508897830 0.588555654080846
0.68468131830753 0.55252750482754 0.589548890521190
0.76097507686579 0.60460997442282 0.593291877272544
0.29485913579866 0.39611113708504 0.589578366850893
0.37278021699982 0.44813152042714 0.588247677318608
0.44804344104077 0.50017098178788 0.588532368930643
0.53064568950433 0.55531752812578 0.588826821961187
0.60720454537870 0.60462058980816 0.588989321597937
0.68515053000372 0.65667142797811 0.589685718183591
0.21780028559570 0.45079619237359 0.587167723857641
0.29521120585179 0.50022919454078 0.588970563297601
0.37286369089539 0.55222545559645 0.588865309856450
0.45090523190190 0.60451037499325 0.588337453890724
0.52910293094584 0.65654311449898 0.588540441605202
0.60690001075275 0.70822402891420 0.588641557802193
0.68261614585767 0.76104140113935 0.587373936295042
0.21729202593140 0.55216254541055 0.590114221653610
0.29527497627235 0.60402509442319 0.590737834455213
0.37310845186328 0.65597781536003 0.589593002717873
0.45115866722690 0.70804226843354 0.589683686518966
0.52898676345219 0.75998919023487 0.590253295868090
0.60682964285144 0.81213967173826 0.588713540913557
0.21797822778340 0.65350243621488 0.594858729467225
0.29511088613090 0.70799189440493 0.592726863072689
0.37338491353052 0.75942902154152 0.590535111360250
0.45110445029968 0.81208794162663 0.591429358953627
0.53053844039806 0.86202229424357 0.593775968501078
0.29622601917486 0.80994775368091 0.590396593953260
0.37407846095974 0.86181780980306 0.591263362623796
0.51289501337995 0.20573383926687 0.588244598767575
0.60269828169534 0.19748933915214 0.588443907398894
0.68067373667584 0.24953868423990 0.590281817649970
0.71925202977503 0.34352564908127 0.591207695453270
0.35610596249025 0.20589962638885 0.589272663696611
0.44421787047128 0.19669133152147 0.589034384144440
0.76032130142609 0.40774084424933 0.590818958398539
0.79733289554173 0.50045934756441 0.591140998812164
0.27774907199434 0.25814875445948 0.592102298816088
0.79712931247777 0.60494875107737 0.594331810130462
0.22684051255782 0.34172963322627 0.592491286450073
0.75995768369368 0.69758643509676 0.592535958937044
0.19965492315927 0.41468113089564 0.586158945983192
0.71877802826006 0.76113269621061 0.586827363567951
0.14848366878799 0.50042665440907 0.587191483544074
0.68005180256631 0.85514375518164 0.586157509906634
0.14856992191736 0.60410316302122 0.592799131997852
0.19982995343619 0.68959145139350 0.596197080783939
0.51251995815625 0.89819782024655 0.595029212782755
0.60222516845509 0.90681726104194 0.591086911314671
0.22703726714333 0.76242174147740 0.591776678959118
0.27789365038122 0.84595582219329 0.589383384575584
0.35620223766842 0.89813027010430 0.590646340331656
0.44427564031559 0.90735732637116 0.590999356436728
0.00000000000000 0.00000000000000 0.370125709308694
0.25000000000000 0.00000000000000 0.370125709308694
0.12500000000000 0.25000002390650 0.370125709308694
0.00000000000000 0.50000004781301 0.370125709308694
0.50000000000000 0.00000000000000 0.370125709308694
0.37500000000000 0.25000002390650 0.370125709308694
0.25000000000000 0.50000004781301 0.370125709308694
0.12500000000000 0.75000007171951 0.370125709308694
0.75000000000000 0.00000000000000 0.370125709308694
0.62500000000000 0.25000002390650 0.370125709308694
0.50000000000000 0.50000004781301 0.370125709308694
0.37500000000000 0.75000007171951 0.370125709308694
0.87500000000000 0.25000002390650 0.370125709308694
0.75000000000000 0.50000004781301 0.370125709308694
0.62500000000000 0.75000007171951 0.370125709308694
0.87500000000000 0.75000007171951 0.370125709308694
0.12500000000001 0.08333334130216 0.370125709308694
0.00000000000001 0.33333336520866 0.370125709308694
0.37500000000001 0.08333334130216 0.370125709308694
0.25000000000001 0.33333336520866 0.370125709308694
0.12500000000001 0.58333338911517 0.370125709308694
0.00000000000001 0.83333341302167 0.370125709308694
0.62500000000001 0.08333334130216 0.370125709308694
0.50000000000001 0.33333336520866 0.370125709308694
0.37500000000001 0.58333338911517 0.370125709308694
0.25000000000001 0.83333341302167 0.370125709308694
0.87500000000001 0.08333334130216 0.370125709308694
0.75000000000001 0.33333336520866 0.370125709308694
0.62500000000001 0.58333338911517 0.370125709308694
0.50000000000001 0.83333341302167 0.370125709308694
0.87500000000001 0.58333338911517 0.370125709308694
0.75000000000001 0.83333341302167 0.370125709308694
0.08082205887734 0.00502254939225 0.302658817823047
0.33082205887734 0.00502254939225 0.302658817823047
0.20582205887734 0.25502257329875 0.302658817823047
0.08082205887734 0.50502259720525 0.302658817823047
0.58082205887734 0.00502254939225 0.302658817823047
0.45582205887734 0.25502257329875 0.302658817823047
0.33082205887734 0.50502259720525 0.302658817823047
0.20582205887734 0.75502262111176 0.302658817823047
0.83082205887734 0.00502254939225 0.302658817823047
0.70582205887734 0.25502257329875 0.302658817823047
0.58082205887734 0.50502259720525 0.302658817823047
0.45582205887734 0.75502262111176 0.302658817823047
0.95582205887734 0.25502257329875 0.302658817823047
0.83082205887734 0.50502259720525 0.302658817823047
0.70582205887734 0.75502262111176 0.302658817823047
0.95582205887734 0.75502262111176 0.302658817823047
0.08835588224530 0.16666668260434 0.302658817823047
0.33835588224530 0.16666668260434 0.302658817823047
0.21335588224530 0.41666670651085 0.302658817823047
0.08835588224530 0.66666673041735 0.302658817823047
0.58835588224530 0.16666668260434 0.302658817823047
0.46335588224530 0.41666670651085 0.302658817823047
0.33835588224530 0.66666673041735 0.302658817823047
0.21335588224530 0.91666675432386 0.302658817823047
0.83835588224530 0.16666668260434 0.302658817823047
0.71335588224530 0.41666670651085 0.302658817823047
0.58835588224530 0.66666673041735 0.302658817823047
0.46335588224530 0.91666675432386 0.302658817823047
0.96335588224530 0.41666670651085 0.302658817823047
0.83835588224530 0.66666673041735 0.302658817823047
0.71335588224530 0.91666675432386 0.302658817823047
0.96335588224530 0.91666675432386 0.302658817823047
0.20582205887736 0.07831079190991 0.302658817823047
0.08082205887736 0.32831081581642 0.302658817823047
0.45582205887736 0.07831079190991 0.302658817823047
0.33082205887736 0.32831081581642 0.302658817823047
0.20582205887736 0.57831083972292 0.302658817823047
0.08082205887736 0.82831086362942 0.302658817823047
0.70582205887736 0.07831079190991 0.302658817823047
0.58082205887736 0.32831081581642 0.302658817823047
0.45582205887736 0.57831083972292 0.302658817823047
0.33082205887736 0.82831086362942 0.302658817823047
0.95582205887736 0.07831079190991 0.302658817823047
0.83082205887736 0.32831081581642 0.302658817823047
0.70582205887736 0.57831083972292 0.302658817823047
0.58082205887736 0.82831086362942 0.302658817823047
0.95582205887736 0.57831083972292 0.302658817823047
0.83082205887736 0.82831086362942 0.302658817823047
0.04417794112264 0.07831079190991 0.437592600794341
0.29417794112264 0.07831079190991 0.437592600794341
0.16917794112264 0.32831081581642 0.437592600794341
0.04417794112264 0.57831083972292 0.437592600794341
0.54417794112264 0.07831079190991 0.437592600794341
0.41917794112264 0.32831081581642 0.437592600794341
0.29417794112264 0.57831083972292 0.437592600794341
0.16917794112264 0.82831086362942 0.437592600794341
0.79417794112264 0.07831079190991 0.437592600794341
0.66917794112264 0.32831081581642 0.437592600794341
0.54417794112264 0.57831083972292 0.437592600794341
0.41917794112264 0.82831086362942 0.437592600794341
0.91917794112264 0.32831081581642 0.437592600794341
0.79417794112264 0.57831083972292 0.437592600794341
0.66917794112264 0.82831086362942 0.437592600794341
0.91917794112264 0.82831086362942 0.437592600794341
0.16164411775470 0.16666668260434 0.437592600794341
0.03664411775470 0.41666670651085 0.437592600794341
0.41164411775470 0.16666668260434 0.437592600794341
0.28664411775470 0.41666670651085 0.437592600794341
0.16164411775470 0.66666673041735 0.437592600794341
0.03664411775470 0.91666675432386 0.437592600794341
0.66164411775470 0.16666668260434 0.437592600794341
0.53664411775470 0.41666670651085 0.437592600794341
0.41164411775470 0.66666673041735 0.437592600794341
0.28664411775470 0.91666675432386 0.437592600794341
0.91164411775470 0.16666668260434 0.437592600794341
0.78664411775470 0.41666670651085 0.437592600794341
0.66164411775470 0.66666673041735 0.437592600794341
0.53664411775470 0.91666675432386 0.437592600794341
0.91164411775470 0.66666673041735 0.437592600794341
0.78664411775470 0.91666675432386 0.437592600794341
0.16917794112266 0.00502254939225 0.437592600794341
0.04417794112266 0.25502257329875 0.437592600794341
0.41917794112266 0.00502254939225 0.437592600794341
0.29417794112266 0.25502257329875 0.437592600794341
0.16917794112266 0.50502259720525 0.437592600794341
0.04417794112266 0.75502262111176 0.437592600794341
0.66917794112266 0.00502254939225 0.437592600794341
0.54417794112266 0.25502257329875 0.437592600794341
0.41917794112266 0.50502259720525 0.437592600794341
0.29417794112266 0.75502262111176 0.437592600794341
0.91917794112266 0.00502254939225 0.437592600794341
0.79417794112266 0.25502257329875 0.437592600794341
0.66917794112266 0.50502259720525 0.437592600794341
0.54417794112266 0.75502262111176 0.437592600794341
0.91917794112266 0.50502259720525 0.437592600794341
0.79417794112266 0.75502262111176 0.437592600794341
\end{verbatim}

\clearpage
The commensurate heterostructure of monolayer hBN containing one $V_B^-$ defect with CrI$_3$ is:
\begin{verbatim}
vb_mono_hbn_crI3                        
1.00000000000000     
    -6.9975797290340003   12.1166494216810001    0.0000000000000000
   -13.9921160724430003   -0.0017570995100000    0.0000000000000000
     0.0000000000000000    0.0000000000000000   23.0000000000000000
B    N    Cr   I 
30    31     8    24
Direct
0.1601934094307174 0.0316975471541642 0.638301065276594
0.1288930039970712 0.2256882188094953 0.638696771165578
0.0968065774357883 0.4193320191042874 0.639028081525040
0.0645556625711941 0.6130051474724018 0.638615597569277
0.0315901586913474 0.8080783324534784 0.638313523858602
0.3223110828501042 0.0646537103094645 0.638636115811134
0.2901921311470587 0.2581272679746919 0.639252275395985
0.2580240781899108 0.4516471401604846 0.639217996226796
0.2255855792538551 0.6454010316312074 0.638678106091418
0.1921148380908772 0.8401907108943304 0.638561929791453
0.4837195390222238 0.0969198336578278 0.639034957266770
0.4515585943070832 0.2902905017570096 0.639238036425698
0.4192301302638092 0.4838275860966392 0.639000892175581
0.3868639496458947 0.6774240026445414 0.638218345825485
0.3545125430248450 0.8710447432782632 0.638326129200652
0.6452849508007817 0.1289952797304523 0.638685619828282
0.6129099364432576 0.3224076046632721 0.638621312470681
0.5806140005387542 0.5160368548336989 0.638768949470336
0.5482826076044128 0.7096552747095622 0.639079455245539
0.5159368154971719 0.9033201716187943 0.638726101713591
0.8079760224065148 0.1602985779996258 0.638298295911872
0.7743057849685244 0.3546204848712080 0.638316476774797
0.7419200961424181 0.5483879766488541 0.639072751420767
0.7095469346566596 0.7420312551220353 0.639051740893557
0.6773061823998364 0.9356953759814516 0.638224218447578
0.9675599585139035 0.1922250575460875 0.638539160074491
0.9355944520834407 0.3869600028260645 0.638212992483220
0.9032071783064313 0.5807185908676620 0.638736371247358
0.8709435440103682 0.7744055542849895 0.638317391112865
0.8400848334749916 0.9676570858213650 0.638547838903384
0.0451781707301544 0.0792125467345304 0.639144914053368
0.0107403997217579 0.2694582197625721 0.638173258391317
0.2046611925632312 0.1078240329635562 0.638210219114064
0.1721181773713685 0.3012258080318665 0.639971569114851
0.1398677089186402 0.4944548680505410 0.638871974434341
0.1077267532995189 0.6874825860635416 0.638213778975660
0.0791024830448705 0.8755817838067141 0.639173383707011
0.3655420662959538 0.1399709028413006 0.638898010501586
0.3332558921484333 0.3333595679539990 0.638914969526508
0.3011219886615152 0.5266245963313260 0.639919914597128
0.2693548412977245 0.7197732287904552 0.638169117267809
0.2375458761889775 0.9136155349344581 0.637776033133889
0.5265290782917434 0.1722199254157269 0.639977294811582
0.4943576781030989 0.3656465070876708 0.638863784010507
0.4624037256272227 0.5589511943421505 0.638361377988314
0.4303649856280099 0.7524546510630976 0.638538508550314
0.3980544891715825 0.9462227206570655 0.638976812513060
0.7196715492258435 0.0108374626340721 0.638155294227618
0.6873838811508299 0.2047577591193259 0.638192561264858
0.6555906410726525 0.3981547532190865 0.638999228850401
0.6236440174425447 0.5914010162284443 0.639323534544338
0.5912937394164866 0.7849237750163113 0.639240259777298
0.5588418077801995 0.9786225305090794 0.638367386590381
0.8754835566117959 0.0452754990271013 0.639146252822399
0.8487084245366260 0.2376429416559806 0.637764261776719
0.8170495511690375 0.4304706574591516 0.638489717282888
0.7848146502014456 0.6237502151059472 0.639271766891241
0.7523460277912997 0.8171572848913581 0.638531436871029
0.9785133451796872 0.4625083999674263 0.638374692995662
0.9461240910951680 0.6556863393054746 0.638952864223640
0.9135136482743909 0.8488100624029311 0.637769558617694
0.0000726468752319 0.5009600843365973 0.420125709308699
0.5011053780870611 0.5008874374613725 0.420125709308699
0.0000000000000000 0.0000000000000000 0.420125709308699
0.5010327312118292 0.9999273531247681 0.420125709308699
0.1671077729042310 0.6679225634903716 0.420125709308699
0.6681405041160673 0.6678499166151468 0.420125709308699
0.1670351260289991 0.1669624791537672 0.420125709308699
0.6680678572408425 0.1668898322785495 0.420125709308699
0.0101869237213990 0.8349320144044086 0.352658817823046
0.5112196549332282 0.8348593675291767 0.352658817823046
0.0101142768461671 0.3339719300678112 0.352658817823046
0.5111470080580105 0.3338992831925935 0.352658817823046
0.3341172238182679 0.6578339617592235 0.352658817823046
0.8351499550301114 0.6577613148839916 0.352658817823046
0.3340445769430502 0.1568738774226190 0.352658817823046
0.8350773081548795 0.1568012305473871 0.352658817823046
0.1570191711730686 0.5110017143075396 0.352658817823046
0.6580519023849121 0.5109290674323219 0.352658817823046
0.1569465242978509 0.0100416299709423 0.352658817823046
0.6579792555096802 0.0099689830957175 0.352658817823046
0.1570661429332887 0.8349107177592074 0.487592600794340
0.6580988741451179 0.8348380708839827 0.487592600794340
0.1569934960580568 0.3339506334226101 0.487592600794340
0.6580262272699002 0.3338779865473782 0.487592600794340
0.3340959271730668 0.5109760391925278 0.487592600794340
0.8351286583849102 0.5109033923173030 0.487592600794340
0.3340232802978420 0.0100159548559304 0.487592600794340
0.8350560115096783 0.0099433079806985 0.487592600794340
0.0101612486063871 0.6578809335194293 0.487592600794340
0.5111939798182235 0.6578082866441974 0.487592600794340
0.0100886017311623 0.1569208491828320 0.487592600794340
0.5111213329429916 0.1568482023076001 0.487592600794340
\end{verbatim}

\clearpage

\section{Effective Exchange Field}
We present the calculated effective exchange field $\textbf{B}_{\text{exc}}$ for all our sampled configuration,
containing 4 distinct stacking twist angles $\theta_{\text{stack}} = 0^\circ, 30^\circ, 60^\circ, 90^\circ$ 
and a $6 \times 6$ grid of lateral translation 
in the CrI$_3$ unit cell. The row and column direction corresponds to fractional
lateral shifts along the (100) and (010) directions of CrI$_3$ unit cell respectively.
Table \ref{tab:1st_flake_angle_0} - Table \ref{tab:1st_flake_angle_90} detail the effective
exchange field $\textbf{B}_{\text{exc}}$ for configurations where the $V_B^-$ defect  is situated in the first hBN layer. Table
\ref{tab:2nd_flake_angle_0} - Table \ref{tab:2nd_flake_angle_90} provide the $\textbf{B}_{\text{exc}}$ values for the instances where the $V_B^-$ defect 
is located in the second hBN layer above the CrI$_3$ substrate.

\begin{table}[h]
    \centering
    \begin{tabular}{|c|c|c|c|c|c|c|}
        \hline
        $\textbf{B}_{\text{exc}}$ [Tesla] & 0 & 1/6 & 1/3 & 1/2 & 2/3 & 5/6 \\
        \hline
        0 & 110.65 & 69.52 & 37.10 & 75.29 & 91.74 & 96.89 \\
        \hline
        1/6 & 63.61 & 97.46 & 78.29 & 116.04 & 122.48 & 73.63 \\
        \hline
        1/3 & 36.83 & 58.84 & 91.73 & 115.25 & 168.24 & 112.96 \\
        \hline
        1/2 & 76.93 & 73.03 & 100.55 & 71.11 & 104.51 & 113.87 \\
        \hline
        2/3 & 90.64 & 97.71 & 111.14 & 71.09 & 32.86 & 72.00 \\
        \hline
        5/6 & 100.25 & 57.44 & 64.22 & 96.95 & 64.45 & 74.93 \\
        \hline
    \end{tabular}
    \caption{Effective exchange field $\textbf{B}_{\text{exc}}$ (in Tesla) for the $V_B^-$ defect located
    in the first hBN layer above CrI$_3$.
    $\theta_{\text{stack}}=0^\circ$}
    \label{tab:1st_flake_angle_0}
\end{table}

\begin{table}[h]
    \centering
    \begin{tabular}{|c|c|c|c|c|c|c|}
        \hline
        $\textbf{B}_{\text{exc}}$ [Tesla] & 0 & 1/6 & 1/3 & 1/2 & 2/3 & 5/6 \\
        \hline
        0 & 101.76 & 71.64 & 71.80 & 101.27 & 95.21 & 74.43 \\
        \hline
        1/6 & 67.93 & 76.84 & 71.76 & 84.49 & 93.43 & 64.77 \\
        \hline
        1/3 & 72.59 & 70.11 & 96.04 & 97.90 & 96.80 & 88.85 \\
        \hline
        1/2 & 102.74 & 61.41 & 78.58 & 100.03 & 83.25 & 98.55 \\
        \hline
        2/3 & 92.63 & 79.48 & 41.23 & 67.12 & 70.48 & 66.33 \\
        \hline
        5/6 & 77.24 & 74.54 & 67.74 & 87.26 & 74.69 & 70.99 \\
        \hline
    \end{tabular}
    \caption{Effective exchange field $\textbf{B}_{\text{exc}}$ (in Tesla) for the $V_B^-$ defect located
    in the first hBN layer above CrI$_3$.
    $\theta_{\text{stack}}=30^\circ$}
    \label{tab:1st_flake_angle_30}
\end{table}

\begin{table}[h]
    \centering
    \begin{tabular}{|c|c|c|c|c|c|c|}
        \hline
        $\textbf{B}_{\text{exc}}$ [Tesla] & 0 & 1/6 & 1/3 & 1/2 & 2/3 & 5/6 \\
        \hline
        0 & 36.07 & 85.78 & 105.72 & 98.35 & 91.48 & 65.53 \\
        \hline
        1/6 & 82.60 & 71.01 & 100.26 & 90.52 & 89.82 & 99.09 \\
        \hline
        1/3 & 108.86 & 89.52 & 92.60 & 89.57 & 68.41 & 95.11 \\
        \hline
        1/2 & 99.89 & 76.66 & 66.55 & 100.88 & 93.86 & 92.86 \\
        \hline
        2/3 & 88.84 & 65.75 & 36.39 & 84.64 & 108.13 & 98.92 \\
        \hline
        5/6 & 64.60 & 96.46 & 83.01 & 69.74 & 90.08 & 75.46 \\
        \hline
    \end{tabular}
    \caption{Effective exchange field $\textbf{B}_{\text{exc}}$ (in Tesla) for the $V_B^-$ defect located
    in the first hBN layer above CrI$_3$.
    $\theta_{\text{stack}}=60^\circ$}
    \label{tab:1st_flake_angle_60}
\end{table}

\begin{table}[h]
    \centering
    \begin{tabular}{|c|c|c|c|c|c|c|}
        \hline
        $\textbf{B}_{\text{exc}}$ [Tesla] & 0 & 1/6 & 1/3 & 1/2 & 2/3 & 5/6 \\
        \hline
        0 & 39.43 & 65.20 & 74.87 & 66.50 & 96.04 & 83.59 \\
        \hline
        1/6 & 64.28 & 78.34 & 103.01 & 88.89 & 97.51 & 102.63 \\
        \hline
        1/3 & 69.03 & 69.80 & 91.47 & 96.07 & 95.95 & 85.54 \\
        \hline
        1/2 & 69.32 & 69.45 & 79.37 & 68.22 & 81.84 & 95.42 \\
        \hline
        2/3 & 96.24 & 74.80 & 100.72 & 74.16 & 72.43 & 98.24 \\
        \hline
        5/6 & 80.44 & 74.75 & 67.00 & 73.77 & 74.79 & 64.73 \\
        \hline
    \end{tabular}
    \caption{Effective exchange field $\textbf{B}_{\text{exc}}$ (in Tesla) for the $V_B^-$ defect located
    in the first hBN layer above CrI$_3$.
    $\theta_{\text{stack}}=90^\circ$}
    \label{tab:1st_flake_angle_90}
\end{table}

\begin{table}[h]
    \centering
    \begin{tabular}{|c|c|c|c|c|c|c|}
        \hline
        $\textbf{B}_{\text{exc}}$ [Tesla] & 0 & 1/6 & 1/3 & 1/2 & 2/3 & 5/6 \\
        \hline
        0 & 12.38 & 9.92 & 8.83 & 12.37 & 16.11 & 14.22 \\
        \hline
        1/6 & 10.02 & 13.07 & 13.27 & 14.76 & 14.77 & 13.27 \\
        \hline
        1/3 & 9.29 & 9.61 & 15.72 & 15.78 & 15.94 & 13.86 \\
        \hline
        1/2 & 13.94 & 10.79 & 13.04 & 12.48 & 14.12 & 14.77 \\
        \hline
        2/3 & 16.34 & 14.42 & 12.31 & 9.60 & 8.10 & 12.12 \\
        \hline
        5/6 & 13.49 & 10.01 & 9.67 & 13.53 & 10.37 & 9.98 \\
        \hline
    \end{tabular}
    \caption{Effective exchange field $\textbf{B}_{\text{exc}}$ (in Tesla) for the $V_B^-$ defect located
    in the second hBN layer above CrI$_3$.
    $\theta_{\text{stack}}=0^\circ$}
    \label{tab:2nd_flake_angle_0}
\end{table}

\begin{table}[h]
    \centering
    \begin{tabular}{|c|c|c|c|c|c|c|}
        \hline
        $\textbf{B}_{\text{exc}}$ [Tesla] & 0 & 1/6 & 1/3 & 1/2 & 2/3 & 5/6 \\
        \hline
        0 & 9.47 & 9.70 & 9.80 & 14.34 & 15.64 & 11.87 \\
        \hline
        1/6 & 8.48 & 11.79 & 15.27 & 12.47 & 13.58 & 13.05 \\
        \hline
        1/3 & 8.90 & 8.55 & 16.17 & 14.88 & 10.10 & 12.71 \\
        \hline
        1/2 & 15.65 & 10.01 & 12.67 & 15.00 & 12.03 & 15.13 \\
        \hline
        2/3 & 16.26 & 12.47 & 10.63 & 11.16 & 9.39 & 13.48 \\
        \hline
        5/6 & 11.45 & 8.93 & 10.27 & 13.25 & 10.41 & 8.81 \\
        \hline
    \end{tabular}
    \caption{Effective exchange field $\textbf{B}_{\text{exc}}$ (in Tesla) for the $V_B^-$ defect located
    in the second hBN layer above CrI$_3$.
    $\theta_{\text{stack}}=30^\circ$}
    \label{tab:2nd_flake_angle_30}
\end{table}

\begin{table}[h]
    \centering
    \begin{tabular}{|c|c|c|c|c|c|c|}
        \hline
        $\textbf{B}_{\text{exc}}$ [Tesla] & 0 & 1/6 & 1/3 & 1/2 & 2/3 & 5/6 \\
        \hline
        0 & 8.43 & 11.33 & 11.55 & 13.71 & 15.39 & 11.98 \\
        \hline
        1/6 & 10.01 & 11.69 & 15.04 & 13.61 & 14.28 & 13.61 \\
        \hline
        1/3 & 11.55 & 10.10 & 15.73 & 16.06 & 15.63 & 13.73 \\
        \hline
        1/2 & 14.85 & 11.03 & 11.59 & 14.19 & 13.66 & 16.15 \\
        \hline
        2/3 & 15.40 & 11.16 & 8.85 & 11.37 & 11.84 & 14.31 \\
        \hline
        5/6 & 11.37 & 9.95 & 10.41 & 12.37 & 11.31 & 11.04 \\
        \hline
    \end{tabular}
    \caption{Effective exchange field $\textbf{B}_{\text{exc}}$ (in Tesla) for the $V_B^-$ defect located
    in the second hBN layer above CrI$_3$.
    $\theta_{\text{stack}}=60^\circ$}
    \label{tab:2nd_flake_angle_60}
\end{table}

\begin{table}[h]
    \centering
    \begin{tabular}{|c|c|c|c|c|c|c|}
        \hline
        $\textbf{B}_{\text{exc}}$ [Tesla] & 0 & 1/6 & 1/3 & 1/2 & 2/3 & 5/6 \\
        \hline
        0 & 10.57 & 10.86 & 9.45 & 13.29 & 16.13 & 12.84 \\
        \hline
        1/6 & 9.60 & 12.27 & 15.23 & 12.55 & 14.70 & 14.57 \\
        \hline
        1/3 & 8.31 & 7.97 & 15.58 & 15.23 & 10.54 & 12.13 \\
        \hline
        1/2 & 13.19 & 8.82 & 11.52 & 14.28 & 12.40 & 13.93 \\
        \hline
        2/3 & 15.50 & 12.22 & 9.98 & 9.58 & 8.90 & 13.85 \\
        \hline
        5/6 & 12.60 & 9.67 & 8.58 & 11.56 & 9.12 & 9.81 \\
        \hline
    \end{tabular}
    \caption{Effective exchange field $\textbf{B}_{\text{exc}}$ (in Tesla) for the $V_B^-$ defect located
    in the second hBN layer above CrI$_3$.
    $\theta_{\text{stack}}=90^\circ$}
    \label{tab:2nd_flake_angle_90}
\end{table}

\clearpage

\bibliography{apssamp}
\bibliographystyle{achemso}